# Robust Data-Driven Linear Power Flow Model with Probability Constrained Worst-Case Errors

Yitong Liu, *Student Member, IEEE,* Zhengshuo Li, *Senior Member, IEEE,* and Junbo Zhao, *Senior Member, IEEE*

*Abstract*—To limit the probability of unacceptable worst-case linearization errors that might yield risks for power system operations, this letter proposes a robust data-driven linear power flow (RD-LPF) model. It is applicable to both transmission and distribution systems and can achieve better robustness than the recent data-driven models. The key idea is to probabilistically constrain the worst-case errors through distributionally robust chance-constrained programming. It also allows guaranteeing the linearization accuracy for a chosen operating point. Comparison results with three recent LPF models demonstrate that the worst-case error of the RD-LPF model is significantly reduced over 2- to 70-fold while reducing the average error. A compromise between computational efficiency and accuracy can be achieved through different ambiguity sets and conversion methods.

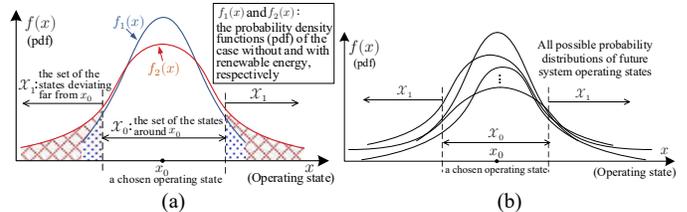

Fig. 1. Illustration of probability distributions of the future system operating state. $\mathcal{X}_1$ denotes the set of states that deviate far from $x_0$, where unacceptable worst-case error may occur. In Fig. 1(a), in the case of high penetration of renewable energy, the pdf of system operating states changes from $f_1(x)$ to $f_2(x)$ where the range of $\mathcal{X}_1$ as well as the occurrence of worst-case errors increases. In Fig. 1(b), since the accurate pdf is an unknown *a prior*, historical operating data can be leveraged to build a data-driven ambiguity set to contain a variety of possible probability distributions.

## I. INTRODUCTION

CONSTRUCTING a linear power flow (LPF) model for a future system operating state is of great interest to the operator, which is useful for various optimization models. LPF can be generally categorized into two types. In the first type, one constructs an LPF model by linearizing the power flow equations with regard to a chosen operating point [1]. In the second type, data-driven techniques, such as partial least squares regression [2],[3], least squares regression [4],[5], support vector regression [6], etc., which utilize historical measurements are employed on top of or in place of the aforementioned linearization for an expected improvement in the linearization accuracy. These data-driven LPF (DD-LPF) models [2]-[6] have demonstrated improved average linearization accuracy as compared to the first-type model.

Nevertheless, an LPF model with moderate average linearization errors but notable worst-case errors still might be improper for certain practical applications. The worst-case errors mean the maximum errors under some system operating states that significantly deviate from the chosen one. For example, it may cause high risks when certain security constraints are on the verge of the limits, where the operators' tolerance of the worst-case error is low. Thus, one should consider reducing both the average error [7] and the worst-case error when constructing an LPF model. Indeed, for the system with increased penetration of variable renewable energy, the future system operating state may notably deviate from the chosen one. In this case, unacceptable worst-case errors are likely to occur, see Fig. 1 for illustrations. However, most DD-LPF models in the literature fail to explicitly bound the worst-case error. Ref. [8] derives the error-bound of the branch LPF model, but the bound obtained with multiple times of inequality relaxations may be relatively loose. Also, [8] does not utilize historical operating data to improve the LPF model accuracy.

To bridge this gap, this letter proposes to construct a robust data-driven LPF model (RD-LPF) whose probability of yielding unacceptable worst-case linearization errors is explicitly constrained through the data-driven distributionally robust chance-constrained (DRCC) programming (DR-CCP). Hence, this RD-LPF model is more robust than existing DD-LPF models. To be more specific, since the future system operating state is unknown, historical operating data are leveraged to build an ambiguity set about the probability distribution of this stochastic future state, over which the unacceptable worst-case errors are probabilistically bounded through chance constraints. The ambiguity set can be moment-based [9] or $\phi$-divergence based [10], etc, depending on the actual preference. In this way, no accurate prior assumption on the distribution of the stochastic future state is required. Also, this RD-LPF model can ensure a guaranteed linearization accuracy for a chosen operating point by considering it as the objective function of the DR-CCP. It is shown that the proposed RD-LPF achieves improved average and worst-case errors for linearization as compared to [4], [5], and [8]. The proposed model is generic and applicable to both transmission and distribution systems.

## II. RD-LPF MODEL

We will first present a generic formulation of the DD-LPF models and its robustified version through the DR-CCP, yielding the proposed RD-LPF model. In this letter, the superscripts T and $-1$ denote transpose and inverse operators, respectively; $|\blacksquare|$ denotes the modulus of a (complex) number.

This work was supported in part by the National Natural Science Foundation of China under Grant 52007105.

Y. Liu and Z. Li are with the School of Electrical Engineering, Shandong University, Jinan 250061, China. Zhengshuo Li is the corresponding author (e-mail: zsli@sdu.edu.cn). J. Zhao is with the Department of Electrical and Computer Engineering, University of Connecticut, Storrs, CT 06269, USA (e-mail: junbo@uconn.edu).

## A. Generic Formulation of DD-LPF Models

Most of the DD-LPF models describe a linear relationship between the dependent variable $\mathbf{y}$ and independent variable $\mathbf{x}$ that can be expressed into the following generic formulation:

$$\mathbf{y} = \mathbf{A}(\mathbf{w})\mathbf{x} + \mathbf{b}(\mathbf{w}), \quad (1)$$

where the specific meanings of $\mathbf{y}$ and $\mathbf{x}$ can be different for different DD-LPF models and they will be elaborated by two examples later; $\mathbf{w}$ is the model parameter vector to be determined; $\mathbf{A}(\mathbf{w})$ and $\mathbf{b}(\mathbf{w})$ denote the matrix and the vector that form the linear relationship between $\mathbf{y}$ and $\mathbf{x}$, and they consist of the parameter $\mathbf{w}$ and other known system parameters.

Given the historical dataset $\{(\mathbf{x}^1,\mathbf{y}^1),...,(\mathbf{x}^K,\mathbf{y}^K)\}$, $\mathbf{w}$ can be obtained by solving the following optimization problem:

$$\mathbf{w} = \arg\min_{\mathbf{w}} \sum_{k}^{K} \sum_{i}^{N_y} r_i(\mathbf{y}^k, \mathbf{x}^k, \mathbf{w}). \quad (2)$$

where $\mathbf{y}^k$ and $\mathbf{x}^k$ denote the $k$-th historical data, respectively; $r_i(\cdot)$ denotes an *error indicator* of measuring the linearization error rated to the $i$-th element of $\mathbf{y}^k$; $K$ and $N_y$ are the numbers of historical data and $r_i(\cdot)$, respectively.

To facilitate understanding of this generic formulation, two DD-LPF models are taken as examples below to explain the specific meaning of the variables in (1) and (2).

The first example is the LSDF model in [4] suitable for transmission systems. To relate it with the above generic model, let $\mathbf{y}$ the vector consisting of the branch active and reactive power flows; let $\mathbf{x}$ the vector consisting of nodal active and reactive power injections; $r_i(\mathbf{y}^k,\mathbf{x}^k,\mathbf{w}) = |y_i^k - \mathbf{a}_i(\mathbf{w})\mathbf{x}^k|^2$, where $y_i^k$ is the $i$-th element of $\mathbf{y}^k$, and $\mathbf{a}_i(\mathbf{w})$ is the $i$-th row of $\mathbf{A}(\mathbf{w})$; each row of $\mathbf{A}(\mathbf{w})$ consists of the elements in $\mathbf{w}$ related to the corresponding row of $\mathbf{y}$, and $\mathbf{b}(\mathbf{w}) = \mathbf{0}$.

Another example is the DD-LPF model in [5], which is applied to distribution systems. Similarly, let $\mathbf{y}$ and $\mathbf{x}$ the vectors consisting of nodal voltages and power injections, respectively; let $\mathbf{w}$ the parameters to be solved by (2) in which $r_i(\mathbf{y}^k,\mathbf{x}^k,\mathbf{w}) = |\varepsilon_i(\mathbf{y}^k,\mathbf{x}^k,\mathbf{w})|$, where $\varepsilon_i(\cdot)$ is the $i$-th row of $\varepsilon(\cdot)$ whose detailed expression can be found in [5]. Moreover, $\mathbf{A}(\mathbf{w})$ and $\mathbf{b}(\mathbf{w})$ can be obtained through simple mathematical operation of nodal admittance matrix and $\mathbf{w}$ [5].

## B. General Idea of Constructing the RD-LPF Model

It can be observed that constructing the generic model through (1) and (2) does not explicitly constrain the magnitude of worst-case linearization errors and their occurrence probability. Hence, the DD-LPF models represented by the above generic model are likely to yield the worst-case errors that may cause risk for system operations. This issue is tackled by constructing an RD-LPF model. The key idea is that the probability of yielding unacceptable worst-case error should be explicitly constrained. To this end, when determining the model parameter $\mathbf{w}$, one needs to modify (2) by adding an explicit probability constraint to ensure that the value of the error indicator not exceeding an acceptable value $\delta$ is credible.

Meanwhile, because the future system state may not emerge equally among all the possible values and is often likely to be close to a certain chosen (e.g., contemporary) operating point, one might be still interested in minimizing the error indicator for this chosen point. This allows satisfying linearization accuracy for this chosen point. It will be achieved by formulating it as the objective function in constructing the RD-LPF model.

## C. Proposed Approach

In the formulation, (3) is used to probabilistically constrain the unacceptable worst-case error for the unknown future system state $\mathbf{x}$. Note that the notation $\boldsymbol{\xi}$ rather than $\mathbf{x}$ is used to represent to the system state because it is deemed stochastic variables in constructing the RD-LPF model.

$$\mathbb{P}\{\tilde{r}_i(\mathbf{y},\boldsymbol{\xi},\mathbf{w}) \leq \delta_i\} \geq 1 - \epsilon_i, i = 1,...,N_y, \quad (3)$$

where $\mathbb{P}\{\cdot\}$ denotes a probability distribution; $\tilde{r}_i(\cdot)$ is a linear equivalent or approximation of $r_i(\cdot)$, for example, for [4]'s $r_i(\cdot)$, $\tilde{r}_i(\cdot) = y_i^k - \mathbf{a}_i(\mathbf{w})\mathbf{x}^k$ then both $\tilde{r}_i(\cdot) \leq \delta_i$ and $\tilde{r}_i(\cdot) \leq -\delta_i$ should be satisfied, and for [5], $\varepsilon_i(\cdot)$ is a complex number and $\tilde{r}_i(\cdot)$ denotes its real part or imaginary part; $\delta_i$ is the upper bound of the acceptable worst-case error; $\epsilon_i$ is the risk level, i.e., the maximum allowed probability of constraint violation.

Constraint (3) indicates that the probability of the equivalent error indicator $\tilde{r}_i$ exceeding the acceptable value $\delta_i$ should be smaller than the risk level. It is often difficult, if not impossible, to accurately presume *a prior* probability distribution function $f(\boldsymbol{\xi})$ of $\boldsymbol{\xi}$, which can be addressed by building an ambiguity set $\mathcal{D}$ that is composed of a set of probability distributions equipped with some common characteristics [10]. Then, (3) can be reformulated to a DRCC as:

$$\inf_{f(\boldsymbol{\xi}) \in \mathcal{D}} \mathbb{P}\{\tilde{r}_i(\mathbf{y},\boldsymbol{\xi},\mathbf{w}) \leq \delta_i\} \geq 1 - \epsilon_i, \quad i = 1,...,N_y. \quad (4)$$

To make (4) tractable, one can construct a moment-based ambiguity set $\mathcal{D}$ and resort to cone duality and convert it to a semi-definite programming (SDP) model as shown in (5) [9]:

$$\gamma_2 \boldsymbol{\Sigma}_0 \cdot \mathbf{G}_i + 1 - l_i + \boldsymbol{\Sigma}_0 \cdot \mathbf{H}_i + \gamma_1 \beta_i \leq \epsilon_i \lambda_i,$$
$$\begin{bmatrix} \mathbf{G}_i & -\boldsymbol{\alpha}_i \\ -\boldsymbol{\alpha}_i^{\mathrm{T}} & 1-l_i \end{bmatrix} \geq \begin{bmatrix} \mathbf{0} & \frac{1}{2}\mathbf{A}_i^w \\ (\frac{1}{2}\mathbf{A}_i^w)^{\mathrm{T}} & \lambda_i + \mathbf{A}_i^x \boldsymbol{\mu}_0 - b_i^x \end{bmatrix}, \quad (5)$$
$$\begin{bmatrix} \mathbf{G}_i & -\boldsymbol{\alpha}_i \\ -\boldsymbol{\alpha}_i^{\mathrm{T}} & 1-l_i \end{bmatrix} \geq \mathbf{0}, \begin{bmatrix} \mathbf{H}_i & \boldsymbol{\alpha}_i \\ \boldsymbol{\alpha}_i^{\mathrm{T}} & \beta_i \end{bmatrix} \geq \mathbf{0}, \lambda_i \geq 0,$$

where $\mathbf{X} \cdot \mathbf{Y}$ denotes the trace of $\mathbf{X}\mathbf{Y}$; $\mathbf{G}_i, l_i, \mathbf{H}_i, \boldsymbol{\alpha}_i, \beta_i$ and $\lambda_i$ denote dual variables of the dual problem; $\mathbf{A}_i^w$ and $b_i^w$ are vectors consisting of $\mathbf{w}$.

The SDP model shown in (5) is equivalent to the original (4) [11]. However, the computational efficiency of the SDP will be low if the scale of the optimal variables is relatively large. In this case, one could also consider using other ambiguity set to speed up the solution while maintaining tractability. To this end, we will introduce a $\phi$-divergence based ambiguity.

Based on the $\phi$-divergence based ambiguity and choosing a reference distribution $\mathbb{P}_0$, the DRCC in (4) is equivalent to a traditional chance constraint [10]:

$$\mathbb{P}_0\{\tilde{r}_i(\mathbf{y},\boldsymbol{\xi},\mathbf{w}) \leq \delta_i\} \geq 1 - \epsilon'_{i_+}, \quad i = 1,...,N_y. \quad (6)$$

where $\epsilon'_{i_+} = \max\{\epsilon'_i, 0\}$. Taking the KL divergence as an example, $\epsilon'_i = \inf_{z \in (0,1)}(e^{-d}z^{1-\epsilon_i} - 1)/(z-1)$, where $z$ is the decision variable, and $d$ denotes the tolerance of the distance between the particular density function and the reference one.

The selection of different ambiguity sets is determined by the actual demand, and we will analyze the accuracy and computational efficiency of the two ambiguity sets in case studies. Furthermore, the error indicator of the chosen operating point could be minimized. Without loss of generality, let the $c$-th historical data be chosen as this operating point, then the DR-CCP is formulated below:

$$\min_{\mathbf{w},\mathbf{G},l,\mathbf{H},\boldsymbol{\alpha},\beta,\lambda} \sum_{i}^{N_y} r_i(\mathbf{y}^c, \boldsymbol{\xi}^c, \mathbf{z}) \quad (7)$$

$$\text{s. t.} \quad (5) \text{ or } (6), \quad i = 1, \ldots, N_y.$$

By obtaining **w** from the SDP model in (5) that can be solved by MATLAB/MOSEK, or from the model in (6) that can be solved by MATLAB/GUROBI, the RD-LPF model is finally constructed in (8). It is suitable for both transmission and distribution systems, which will be verified in case studies.

$$\mathbf{y} = \mathbf{A}(\mathbf{w})\mathbf{x} + \mathbf{b}(\mathbf{w}), \text{ where } \mathbf{w} \text{ is solved from (7)}. \quad (8)$$

### III. CASE STUDIES

The linearization error of our model is compared with the DD-LPF models in [4], [5], and the error-bound model in [8]. Since the models in [4] and [8] are mainly for transmission systems, the IEEE 118-bus and Polish 2383-bus transmission systems from the MATPOWER are used for comparisons. As the model in [5] is for distribution systems, the IEEE 123-node and 8500-node three-phase distribution systems from OpenDSS are used for comparison. The historical data are obtained by randomly changing the net load level within ±20% of the base net load level, where the net load fluctuations consist of those from renewable energy and loads. The linearization errors are measured in terms of relative errors, where the exact values of the power flow are obtained by the power flow engines in MATPOWER and OpenDSS.

Tables I and II show the average and worst-case errors of our model and the others for the net load levels being 60%, 80%, 120%, and 140% of the base net load level. Our M1 and M2 approaches denote the models, where **w** is solved from (7) with (5) and (6) as constraints, respectively. The results indicate that compared with the DD-LPF models in [4] and [5], both the average and the worst-case errors of our two models are notably smaller, e.g., reduced 2- to 3-fold. Table I also indicates that as compared to the error-bound model in [8], the worst-case error of our model is reduced by approximately 10- to 70-fold. The relatively large worst-case error of [8]'s model might be due to the loose error-bound as well as not leveraging historical data to improve the linearization accuracy.

Note that (5) is an SDP whose solution time is significantly affected by the number of decision variables determined by the forms of $r_i(\cdot)$. For example, in [5], the average solution time of both test systems for each node is under $2\times10^{-3}$s. However, the number of decision variables for [4] are related to the system scale and the average solution time of the 118-bus system is about 40s. This is about tens of minutes for the 2383-bus system. The computational efficiency can be enhanced if the $\phi$-divergence is used to describe the ambiguity set, with which the average solution time of the 2383-bus system is reduced to tens of seconds. Though Tables I indicates that the $\phi$-divergence based model may be relatively conservative, the linearization error is still more than 2-fold smaller than the other two models. In fact, the proposed M2 model only has slightly degraded performance while achieving significantly improved computational efficiency as compared to that of the M1 model. This means that one can choose which kind of ambiguity set to use according to the actual demand for balancing the solution accuracy and computational efficiency. Our proposed model is flexible enough to offer that.

### IV. CONCLUSION

This letter proposes constructing an RD-LPF model through DR-CCP that explicitly constrains the probability of unacceptable worst-case errors based on two kinds of ambiguity sets. This model is suitable for both transmission systems and distribution systems. Case studies confirm that compared with the recent LPF models [4],[5],[8], the RD-LPF model's worst-case errors can be reduced over 2- to 70-fold and its average error is also reduced. Future works will focus on testing more efficient computing algorithms, e.g., parallel alternating direction method of multipliers-based methods, to deal with constraints in the form of SDP in (7) while maintaining high computational efficiency.

TABLE I
ERRORS OF THE OUR MODEL AND [4]'S AND [8]'S MODELS (UNITS: ×10$^{-3}$ P.U.)

| Load level | | 118-bus system | | | | 2383-bus system | | | |
|---|---|---|---|---|---|---|---|---|---|
| | | Our M1 | Our M2 | [4]'s model | [8]'s model | Our M1 | Our M2 | [4]'s model | [8]'s model |
| Avg. | 60% | 0.88 | 1.21 | 1.79 | 2.91 | 1.97 | 2.54 | 3.18 | 8.35 |
| | 80% | 0.37 | 0.40 | 0.68 | 2.54 | 0.40 | 0.49 | 0.67 | 7.41 |
| | 120% | 0.29 | 0.33 | 0.40 | 1.92 | 0.86 | 0.94 | 1.47 | 9.75 |
| | 140% | 0.97 | 1.32 | 1.94 | 1.73 | 3.43 | 4.11 | 7.62 | 10.9 |
| WC. | 60% | 6.70 | 8.03 | 13.2 | 73.1 | 3.91 | 4.37 | 7.24 | 18.5 |
| | 80% | 0.88 | 0.95 | 1.82 | 73.7 | 0.85 | 0.96 | 1.97 | 20.8 |
| | 120% | 0.87 | 0.93 | 2.17 | 58.4 | 2.01 | 2.38 | 4.16 | 45.2 |
| | 140% | 4.18 | 4.96 | 9.99 | 50.9 | 6.96 | 7.78 | 14.85 | 59.2 |

TABLE II
ERRORS OF THE OUR MODEL AND [5]'S MODEL (UNITS: ×10$^{-3}$ P.U.)

| Load level | | IEEE 123-node system | | | IEEE 8500-node system | | |
|---|---|---|---|---|---|---|---|
| | | Our M1 | Our M2 | [5]'s model | Our M1 | Our M2 | [5]'s model |
| Avg | 60% | 0.55 | 0.59 | 1.00 | 0.61 | 0.68 | 1.01 |
| | 80% | 0.68 | 0.83 | 1.24 | 0.65 | 0.76 | 1.09 |
| | 120% | 0.93 | 0.98 | 1.33 | 1.01 | 1.14 | 1.57 |
| | 140% | 1.15 | 1.46 | 2.52 | 1.22 | 1.57 | 3.06 |
| WC. | 60% | 0.89 | 0.97 | 1.95 | 0.94 | 1.07 | 2.13 |
| | 80% | 1.57 | 1.81 | 2.80 | 1.49 | 1.65 | 2.51 |
| | 120% | 1.65 | 1.86 | 3.00 | 1.58 | 1.73 | 2.79 |
| | 140% | 2.84 | 3.01 | 5.98 | 2.96 | 3.22 | 6.02 |